\documentclass{jfm}

\usepackage{graphicx}

\usepackage{newtxtext}
\usepackage{newtxmath}
\usepackage{natbib}
\usepackage{hyperref}
\hypersetup{
    colorlinks = true,
    urlcolor   = blue,
    citecolor  = black,
}

\usepackage{float}
\usepackage[dvipsnames]{xcolor}
\usepackage[justification=justified]{caption}
\usepackage{textgreek}
\definecolor{blueink}{rgb}{0, 0, 0.8}

\newcommand{\RomanNumeralCaps}[1]

\title{Sliding dynamics of a particle in a soap film}

\author{Youna Louyer \aff{1},
  Benjamin Dollet \aff{2},
  Isabelle Cantat \aff{1}
 \and Anaïs Gauthier\aff{1}
   \corresp{\email{anais.gauthier@univ-rennes.fr}}}

\affiliation{\aff{1}Univ Rennes, CNRS, IPR (Institut de Physique de Rennes) - UMR 6251, F- 35000 Rennes
\aff{2}Université Grenoble Alpes, CNRS, LIPhy, 38000 Grenoble, France}

\begin{document}

\maketitle

\begin{abstract}

We investigate the sliding dynamics of a millimeter-sized particle trapped in a horizontal soap film. Once released, the particle moves toward the center of the film in damped oscillations. We study experimentally and model the forces acting on the particle, and evidence the key role of the mass of the film on its shape and particle dynamics. Not only is the gravitational distortion of the film measurable, it completely determines the force responsible for the motion of the particle -- the catenoid-like deformation induced by the particle has negligible effect on the dynamics. Surprisingly, this is expected for all film sizes as long as the particle radius remains much smaller than the film width. We also measure the friction force, and show that ambient air and the film contribute almost equally to the friction. The theoretical model that we propose predicts exactly the friction coefficient as long as inertial effects can be neglected in air (for the smallest and slowest particles). The fit between theory and experiments sets an upper boundary $\eta_s \leq 10^{-8}$ Pa\,s\,m for the surface viscosity, in excellent agreement with recent interfacial microrheology measurements.
\end{abstract}


\section{Introduction}
\label{Introduction}

Despite their ephemeral nature and apparent fragility, soap bubbles and soap films can withstand large stresses. As first shown by \cite{Courbin:2006b} and later by \cite{Gilet:2009a}, soap films act as non-linear vertical springs (liquid trampolines) capable of repelling a drop or a particle approaching at low velocity. For more violent impacts, they deform so much that they let the incoming object pass without rupturing \citep{Pan:2007, Fell:2013}. They capture instead a fraction of the particle kinetic energy \citep{LeGoff:2008}, making foams capable of stopping projectiles. The transition between bouncing and crossing is governed by the Weber number, which compares the surface energy of the film to the kinetic energy of the particle \citep{Gilet:2009a}. Using this criterion, soap films can be used as liquid sieves, trapping the small and slow objects while allowing the larger and faster ones to pass \citep{Stogin:2018}.

Soap films are usually modelled quasi-statically, considering only the effect of surface tension forces: they are often considered and studied as real life minimal surfaces \citep{Almgren:1976, Courant:1940, Goldstein:2010}. In presence of a drop or a particle, the expected minimal surface is a catenoid, a geometry that matches relatively well the shape of a film deformed by an impacting object \citep{Gilet:2009a, Chen:2019}. Experimentally, the weight of the liquid contained in a soap film is not expected to significantly affect its geometry as long as the film dimension does not exceed a characteristic length $\ell \sim {\gamma}/{(\rho g e)}$, where $\gamma$ is the surface tension of the film, $e$ its thickness, $\rho$ the density of the liquid and $g$ gravity \citep{Cohen:2017}. For a film of thickness $e$ = 10 \textmugreek m, the length $\ell$ is equal to $30$ cm, which is larger than the films usually studied. Here, we consider the dynamics of a millimeter-sized marble trapped in a 10 centimeter-wide horizontal soap film. For this problem, we evidence in contrast the key role of the intrinsic weight of the film, which deformation is at the origin of the particle motion. We also focus on the drag force experienced by the bead and evidence the almost equal contributions of air and the film. 

\section {Experiment}
\label{Experiment}

\begin{figure}
\centering
\includegraphics[width=0.8\textwidth]{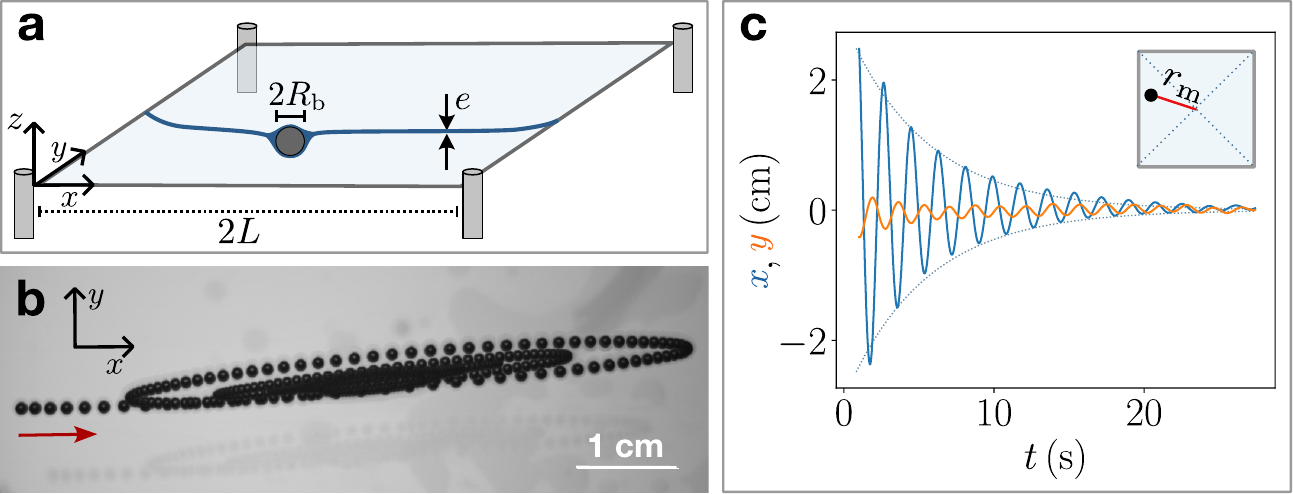}
\caption{\textbf{a.} A bead with radius $R_{\rm b}$ is deposited in a square soap film of size 2L $\sim$ 10 cm and thickness $e \sim$ 10 \textmugreek m. \textbf{b.} Top view. Chronophotography showing the trajectory of a bead  of mass $m_{\rm b}$ = 3.7 mg (effective mass $m$ = 4.41 mg) and radius $R_b$ = 0.5 mm within the film, as seen from the top. Two different images of the particle are separated by 20 ms. \textbf{c.} Position ($x,y$) of the particle relative to the center of the film as a function of time $t$. It follows closely what is expected from a damped harmonic oscillator, with a pseudo-pulsation $\omega = 3.4 \pm 0.1$ s$^{-1}$, and an exponentially decaying envelop (dotted line) $A\,e^{-t/\tau}$, with $A = 2.9 \pm 0.1$~cm and $\tau = 5.4 \pm 0.2$ s.}
\label{figure1}
 \end{figure}

A soap film with dimensions $2L \times 2L$ (with 3.3 cm $\leq L \leq$ 5 cm) is produced by dipping a square frame in a soap solution consisting of 5.6 g/L of sodium dodecyl sulfate (SDS), 50 mg/L of dodecanol in a water-glycerol mixture (15\% of glycerol in volume). The concentration of SDS is more than two times the critical micellar concentration (CMC$_{\rm SDS}$ = 2.37 g/L), and the surface tension of the film is equal to $\gamma$ = 33.2 $\pm$ 0.1\,mN/m. Glycerol is used to reduce evaporation so that a film lasts 1 to 3 minutes before rupture. To ensure the repeatability of the experiments, the frame is removed from the bath at a constant velocity $V_{\rm motor}$ using a motorized stage. The thickness $e$ of the film is calibrated as a function of $V_{\rm motor}$, by puncturing the film and following the growth of the hole with a high speed camera (Phantom Miro LAB3a10) at 6000 fps. The opening velocity $V$ is related to $e$ by the Taylor-Culick law: $e = 2 \gamma/\rho V^2$ \citep{Taylor:1959, Culick:1960}, with $\rho \simeq 1042$ kg/m$^3$ the density of the soap solution. $e$ is thus varied in a controlled manner between 3 and 25 \textmugreek m, with an error of 15\%. In almost all experiments, $V_{\rm motor}$ = 20 cm/s and a film thickness $e$ = $9.8 \pm 1.4$ \textmugreek m is expected. Approximately 10~s after the film fabrication, a millimeter-sized particle (Silibeads from Sigmund Lindner) with radius $R_b$ (250 \textmugreek m $\leq R_{\rm b} \leq$ 750 \textmugreek m), mass $m_{\rm b}$ and density $\rho_b$ (2580 kg/m$^3$ $\leq \rho_b \leq$ 9200 kg/m$^3$) is deposited in the film, a few centimeters from the center (Fig \ref{figure1}a). The particle is initially wet by the soap solution, so that it is held \textit{in} the film after its release. It is thus surrounded by a meniscus which height and width increase with time as the liquid in the film is drained towards the bead by capillary suction \citep{Aradian:2001, Guo:2019}. For a given extension $R$ of the meniscus (relative to the center of the particle), the shape of the meniscus is fully determined using the model of \cite{Orr:1975} (Figure \ref{figureA1}a). The volume and mass of the meniscus can thus be exactly calculated.

\medskip

Figure \ref{figure1}b illustrates the motion of a particle with radius $R_b$ = 0.5 mm and mass $m_b = 3.7$~mg after its release in the film without initial velocity. The bead spontaneously slides towards the center of the film, following a very flattened spiral trajectory (see also Movie 1). The positions $x$ and $y$ of the particle, relative to the film center, are plotted in blue and orange in Figure~\ref{figure1}c as a function of time~$t$. They exhibit damped harmonic oscillations with a nearly identical period but different amplitudes, due to the nature of the trajectory. The position of the particle along the primary direction of the spiral is fitted by a damped sinusoidal function of the form $A \exp(-t/\tau)\sin(\omega t+\phi)$, with $A$ and $\phi$ two constants. In Figure~\ref{figure1}c, the pseudo-pulsation is $\omega = 3.4 \pm 0.1$ s$^{-1}$ and the characteristic time of the exponential envelope (shown with a dotted line) is $\tau = 5.4 \pm 0.2$ s. Damped oscillations are observed for all the particles that we tested (see for example Movies 2 and 3). Their presence indicate that, at the dominant order, the particle is \textit{i)} driven by a spring-like force $\boldsymbol{F} = -k(\boldsymbol{x+y}) = -k\boldsymbol{r_{\rm m}}$ (with $k$ the spring constant and $\boldsymbol{r_{\rm m}}$ the radial distance between the bead and the film centers) and \textit{ii)} slowed down by a viscous drag force $\boldsymbol{F}_{\rm friction} = -\alpha \boldsymbol{\varv}$ (with $\alpha$ the friction coefficient and $\boldsymbol{\varv}$ the particle velocity). The friction coefficient $\alpha = {2m}/{\tau}$ and the spring constant $k = m\left(\omega^2 + 1/\tau^2\right)$ are deduced from $\omega$ and $\tau$ using a damped harmonic oscillator model. $m$ is here the effective mass of the moving object, taking into account the added mass of the meniscus around the particle.

With the reflective lighting used here (where the soap film acts as a mirror reflecting light to the camera), the meniscus appears as a darker area surrounding the particle in the top-view images. Its size $R$ increases over the course of the oscillations; from $R = 2.05\,R_{\rm b}$ (relative to the center of the particle) at $t = 0$ to $3.63\,R_{\rm b}$ after 30 seconds in Fig. \ref{figure1}b. The curve $R(t)$ is presented in Figure \ref{figureA1}b. The density of the particles being significantly higher than that of the film, the meniscus growth causes a modest variation of the effective mass $m$ with time: in Fig. \ref{figure1}b (and Figure \ref{figureA1}b), $m$ varies from $1.08\,m_{\rm b}$ at $t = 0$ to $1.24\,m_{\rm b}$ at t = 30 s. Experimentally, $\alpha$ and $k$ are calculated from a fit over the duration of the oscillations: the value obtained is therefore a time average between the start and end of the oscillations. For this reason, we also use an average effective mass $m$ for the moving object. In practise, $m$ is measured at a time $t = 2.5 \tau$, corresponding to half of the oscillation duration (green dotted line in Figure \ref{figureA1}b). As shown in Figure \ref{figureA1}c, $m$ is typically 20\% higher than the mass $m_{\rm b}$ of the particle alone.

\noindent In the rest of the manuscript, we systematically take into account the added mass of the meniscus and use the effective mass $m$ for the particle.

\smallskip
\noindent To model the particle dynamics, we first study in section~\ref{Film} the film deformation, which is then used in section \ref{SpringForce} to model the spring force. In section \ref{Friction}, we finally characterize the friction force.


\section{Film deformation by a static particle}
\label{Film}

\begin{figure}
\begin{center}
\includegraphics[width=0.9\textwidth]{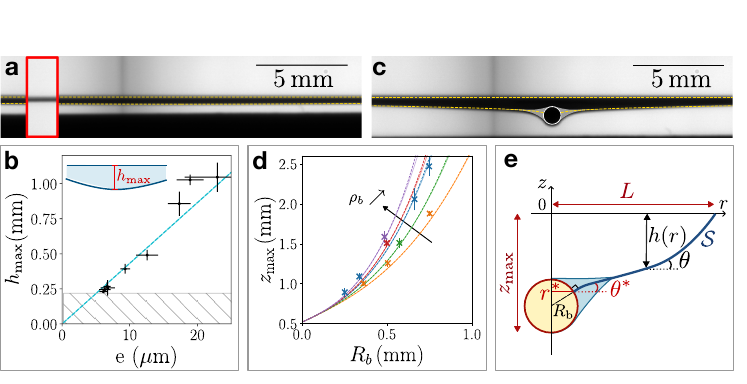}
\caption{\textbf{a.} Side-view image of a soap film attached to a frame of size $2L$ = 6.7 cm, evidencing the film deformation under its weight. The dotted line is the full numerical solution of the film shape. The inset (in red) is a picture of the frame without the film. \textbf{b.} Maximum deformation $h_{\rm max}$ of the film as a function of its thickness $e$. The hatched area is a region where $h_{\rm max}$ cannot be measured. Eq. \ref{model_hmax} is shown with a continuous line, the numerical solution is a perfectly superposed dotted line. \textbf{c.} Deformation of the film in presence of a marble of radius $R_{\rm b}$ = 0.5 mm (with theoretical position shown with a white circle). The dotted line is the numerical solution of the film shape, the meniscus shape is added in grey. \textbf{d.} Maximum deformation $z_{\rm max}$ as a function of $R_{\rm b}$. The experiments are shown with crosses, for varying $\rho_{\rm b}$: 2580 kg/m$^3$ (orange), 4100 kg/m$^3$ (green), 6300 kg/m$^3$ (blue), 7100 kg/m$^3$ (red) and 9200 kg/m$^3$ (purple) and compared with  Eq. \ref{model_zmax} (continuous lines). The numerical solution is shown with a dotted line. The best fit is obtained for $e = 12$~\textmugreek m. \textbf{e}. Notations used to model the film shape.}
\label{figure2}
\end{center}
\end{figure}

The film deformation is observed from the side using a back-light illumination. The light source (a square LED light of width 50 cm and height 25 cm) is placed 2 m away from the film, and a vertical plate with a slit of height 1 cm is positioned a few centimetres in front of the frame. This setup ensures that the incident light rays on the film are almost parallel, with a variation of incident angles smaller than 3 degrees. The soap film, consisting of two parallel interfaces, does not deflect the light. However, a large majority of the light intensity is reflected when the film is illuminated by a grazing light: a darker area is then visible on the camera sensor, corresponding to the projection of the film shape in the $(y,z)$ plane. The frame used here is a nylon wire of diameter 120 \textmugreek m held between four vertical posts. It appears as a fuzzy gray line of width 220 \textmugreek m. It is  shown (in absence of a film) in the inset of Figure \ref{figure2}a (red rectangle), with the same scale and position as the rest of the picture.

Figure \ref{figure2}a shows an image of a film of size $2L$ = 6.7 cm obtained with this method (the black area below the film is the edge of the slit). The dark region induced by the reflection of the ray lights on the film is $393\pm35$ \textmugreek m thick: it is much larger than the film thickness ($e$ = 10 \textmugreek m) or the apparent frame diameter (red inset), indicating that the film is deflected. In Figure \ref{figure2}b the maximum deflection of the film $h_{\rm max}$ is measured for various film thicknesses $e$, obtained by puncturing the film. The error bars show the standard deviation of ten experiments, and the hatched area indicates the region where the film shadow is smaller than the apparent diameter of the frame. The amplitude of the deflection $h_{\rm max}$ increases linearly with $e$; it reaches 1 mm for the thicker films ($e$ = 23 \textmugreek m). 
Figure \ref{figure2}c is a picture of the film of Fig. \ref{figure2}a, now holding a marble of radius $R_{\rm b} = 0.5$~mm and density $\rho_b$ = 9200 kg/m$^3$ (which theoretical position is highlighted by a white circle). The marble induces an additional deformation of the order of the initial film deflection. To characterize it, we plot in Figure \ref{figure2}d the maximum deflection $z_{\rm max}$ measured as the distance between the top of the film and the bottom of the marble. We vary both the bead radius $R_{\rm b}$ and its density $\rho_b$. The crosses are the experimental measurements, and the dotted lines show equation \ref{model_zmax}. $z_{\rm max}$ increases both with $R_{\rm b}$ and $\rho_b$, with an offset of 530 \textmugreek m for $R_{\rm b} = 0$ associated with the film deformation in absence of a particle. 

\smallskip

The film geometry is determined using an axisymmetric model, thus assuming a circular frame. The film height is noted $h(r)$ with $r$ the distance to the center of the film in cylindrical coordinates, in the ($x,y$) plane of the frame. The film is identified with the surface ${\cal S}$ given by $z=h(r)$ (see Figure \ref{figure2}e for a definition of the variables). Close to the bead, the two film interfaces separate to form a meniscus: in that region, $\cal{S}$ (shown as a dark blue line in Figure \ref{figure2}e) is the mid surface between the two interfaces. As the particle is entirely wet, ${\cal S}$ is perpendicular to the solid along the contact line, located at  $r = r^* = R_{\rm b}\cos\theta^*$, with $\theta^*$ the angle between the normal to the particle and the horizontal at the contact line. $\theta^*$ is given by the vertical equilibrium between the surface tension force and the particle's weight: $2 \times 2 \pi \gamma r^* \sin\theta^* = mg$ so that $\sin(2\theta^*) = {mg}/{(2\pi\gamma R_{\rm b})}$. The condition $\sin(2\theta^*) \leq 1$ sets the criteria at which a particle of mass $m$ (density $\rho_b$) and radius $R_{\rm b}$ can be held within the soap film. For $\rho_{\rm b} = 6300$ kg/m$^3$, the critical bead radius is $R_{\rm c,th} = 0.90$ mm. This is consistent with our experimental observations, where 0.75 $<R_c<$ 1.05 mm.

\smallskip

$h(r)$ is determined by the force balance on a film element, projected on its  normal, where the Laplace pressure due to the two curved liquid interfaces $2\gamma \kappa$ (with $\kappa$ the local curvature of the film) balances that of the weight of the film $\rho g e \cos\theta$ (with $\theta$ the angle between the tangent to the film and the horizontal, or equivalently between the normal to the film and the horizontal). 
The equation $\kappa = \rho g e \cos\theta/(2\gamma)$ is solved numerically using the same parametrization as \cite{Cohen:2017}, and the predicted film shape is shown with dotted lines in Figures \ref{figure2} a and c. It matches well the film distortion, with the film thickness as an adjustable parameter. The best fit is obtained for $e = 11$~\textmugreek m, which corresponds to the film calibration: $e = 9.8 \pm 1.4$~\textmugreek m. In our experiments, $\|\bnabla h\| \sim h_{\rm max}/L \simeq 10^{-2}~\ll~1$, meaning that the problem can be simplified to a small deflection situation and solved analytically. The curvature of the film is then $\kappa =  \Delta h$ and $\cos\theta \simeq 1$ so that $h(r)$ is the solution of:  
\begin{align}
 \Delta h = \frac{\rho g e}{2\gamma}.
    \label{equation_film_shape}
\end{align}
\noindent For a particle placed at the center of the film, Equation \ref{equation_film_shape} is solved with the following constraints: \textit{i)} the film is attached to the frame, so that $h = 0$ in $r = L$ and \textit{ii)} it is attached to the particle in $r = r^* = R_{\rm b}\cos\theta^* = R_{\rm b}$. For small deflections, the attachment condition simplifies to 
${\mathrm{d} h}/{\mathrm{d} r}|_{r = R_{\rm b}} = \theta^* = {mg}/{(4 \pi \gamma R_{\rm b})}$. The physical boundary condition of a $\pi/2$ contact angle between the mid plane of the film and the bead is replaced in this limit by a condition of contact along the bead diameter at a free angle; an approximation that we will keep in the following models. The film shape is then:
\begin{align}
h(r) &= \frac{mg}{4\pi \gamma} \ln\frac{r}{L} + \frac{\rho g e}{8 \gamma}(r^2 - L^2) ,
    \label{film_shape}
\end{align}
\noindent with the approximation that $\rho \pi R_{\rm b}^2 e \ll m$. Due to the linearity of Eq. \ref{equation_film_shape}, $h(r)$ is the sum of a logarithmic deformation caused by the particle mass and a parabolic deformation due to the film mass. The maximum deflection of the film is then directly deduced from \ref{film_shape}: 
\begin{align}
&h_{\rm max} = \max(|h|) = \frac{\rho g e L^2}{8 \gamma} \text{ in absence of the bead} 
\label{model_hmax}
\\
\text{and } &z_{\rm max} = \frac{mg}{4 \pi \gamma} \ln\frac{r}{L} + \frac{\rho g e}{8 \gamma}(L^2 - R_{\rm b}^2) + R_{\rm b}\left (1+ \frac{mg}{4\pi\gamma R_{\rm b}}\right) \text{ with the bead.
}
\label{model_zmax}
\end{align}

\noindent Equation \ref{model_hmax} is shown with a continuous line in figure \ref{figure2}b: the small deflection approximation perfectly overlaps the complete numerical solution of the problem (dotted line). It also fits the experimental measurements without adjustable parameter. A good fit between Eq. \ref{model_zmax} (continuous line) and the experiments is also observed in figure \ref{figure2}d. A small difference between the linearized solution and the complete numerical solution (dotted lines) is visible for the largest $R_{\rm b}$, but it remains much smaller than the experimental error bars. Here, the best fitting parameter is $e$ = 12 \textmugreek m, a value close to the calibration of the film thickness.

\section{Film force on a bead}
\label{SpringForce}

The force exerted by the film on the bead can be decomposed into two components. The vertical component, $F_{\perp}$, holds the particle in the film by counterbalancing its weight: $F_{\perp} = mg$. The in-plane component, $F$ drives the motion of the particle (Figure \ref{figure1}b). In the following, we focus on the characteristics of the in-plane force, $F$.

\smallskip

\begin{figure}
\begin{center}
-\includegraphics[width=0.9\textwidth]{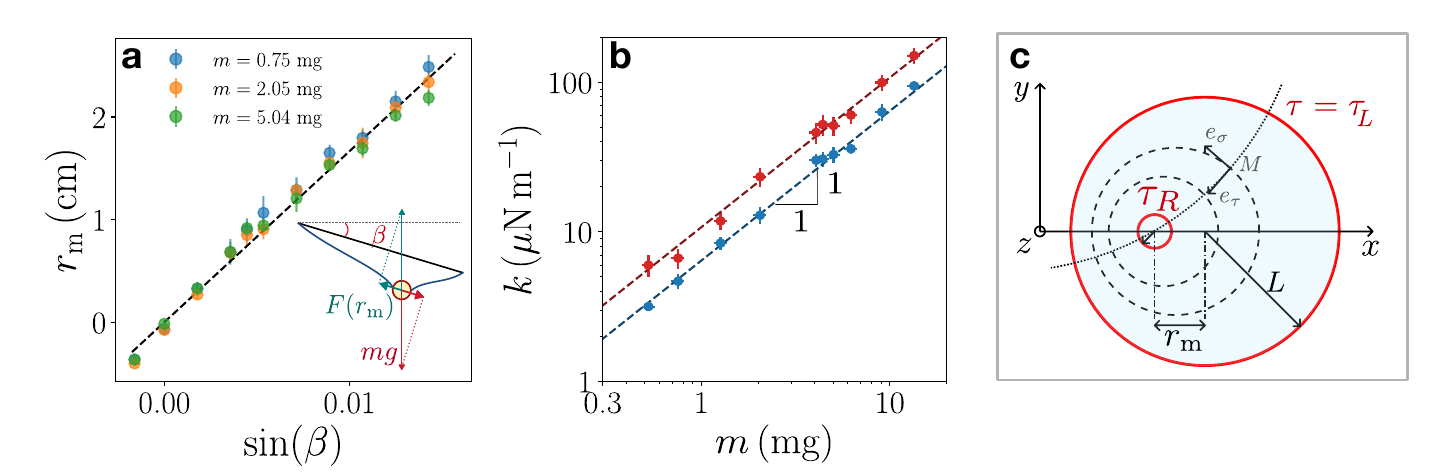}
\caption{\textbf{a.} Equilibrium position $r_{\rm m}$ of the particle as a function of the tilt angle $\sin\beta$. The experiments are shown with dots (blue: $m=0.75 $ mg, orange: $m=2.05$ mg and green: $m=5.04$ mg); the error bars are the standard deviation of 10 measurements. The dotted line show Eq. \ref{eq_F1}, with fitting parameter $e=8.0 $ \textmugreek m. The inset is a schematic of the tilted frame experiment. \textbf{b.} Spring constant $k$ as a function of the effective mass $m$ (static measurements in blue, dynamic measurements in red). Eq. \ref{eq_F1} is shown with dotted lines, with fitting parameter $e = 8.9$ µm for the static experiment (blue) and $e = 14.2$ µm for the dynamic experiment (red). In both cases, the fitting parameter matches the film thickness calibration (two different frames are used). \textbf{c.} In bipolar coordinates, any point M of the plan has coordinates $(\sigma,\tau,z)$ with an orthonormal basis $(\mathbf{e}_\sigma,\mathbf{e}_\tau,\mathbf{e}_z)$. The iso-$\tau$ curves are non-intersecting circles: the circular frame is defined by $\tau = \tau_L$ and the equator of the particle is $\tau = \tau_R$ (both shown in red).}
\label{figure3}
\end{center}
\end{figure}

\noindent $F$ is first measured in static conditions, by tilting the frame at an angle $\beta$. $\beta$ is kept small enough (below one degree) that, in the absence of a particle, the film retains its parabolic shape. When the frame is tilted, the particle stabilises at a position $r_{\rm m}$ where the film force $F$ balances the projection of the weight of the particle in the ($x,y$) plane  $m g \sin\beta$ (see the inset of Fig. \ref{figure3}a). In Figure \ref{figure3}a, the equilibrium position $r_{\rm m}$ is plotted as a function of the tilt angle $\sin\beta$ for three beads with effective mass $m$ = 0.75 mg (blue), $m$ = 2.05 mg (orange) and $m$ = 5.04 mg (green). The equilibrium position of the particle does not depend on $m$, as evidenced by the collapse of the data. In addition, $r_{\rm m}$ increases linearly with $\sin\beta$, which indicates that the in-plane film force on the bead, $F = mg\sin\beta$ is linear in the domain where it is measured (75\% of the width of the frame). This is consistent with the particle dynamics in an horizontal frame, where a linear spring force $F \sim k r_{\rm m}$ is deduced from damped harmonic oscillations. In Figure \ref{figure3}b, the spring coefficient $k$ is plotted as a function of the effective mass $m$. Here, $k$ is determined from tilting the film (blue dots) and from the particle oscillations (red dots). In both cases, $k$ increases linearly with $m$ (which is varied by a factor $\simeq$ 20 between 0.5 and 13.4 mg). Note that two different frames are used in these experiments (a nylon wire frame of size $2L = 6.7$ cm for the blue points and a 2 mm thick frame with $2L = 10$ cm for the red points), so that the exact values of $k$ cannot be directly compared.

\smallskip

To model this spring force, the film shape equation \ref{equation_film_shape} has to be solved for a particle off-centered by a distance $r_{\rm m}$. Due to the linearity of (\ref{equation_film_shape}), the film deformation $h$ is the superposition of \textit{i)} the deformation caused by the weight of the film $h_{\rm 1} = \rho ge(r^2 - L^2)/8\gamma$, as previously determined (Eq. \ref{film_shape}) and \textit{ii)} the deformation $h_{\rm 2}$ of a weightless film subjected only to the weight of an off-centered particle. We start by calculating the force that the deformation $h_{\rm 1}$ alone would create. In this limit, the film shape does not depend on the bead position and its surface energy is constant. The particle is held by the film at a height $z = h_{\rm 1}(r_{\rm m})$: it is thus in a potential well of equation $E_{\rm} = mgz$. This produces an horizontal spring force of amplitude $F_{\rm 1} = \lvert - \partial E_{\rm}/\partial r_m \rvert$: 
\begin{align}
 F_{\rm 1} = \frac{m \rho e g^2}{4 \gamma} r_{\rm m} .
\label{eq_F1}
\end{align}

\noindent Interestingly, the force $F_{\rm 1}$ alone reproduces remarkably well the experimental data. First, similarly to what is expected of a particle in a tilted bowl, the equilibrium position of the bead varies linearly with the tilt angle $\sin \beta$ and does not depend on the mass of the particle. In Figure \ref{figure3}a, $r_m$ = f($\sin \beta$) is fitted by a linear curve of slope $4\gamma/\rho g e$, as expected from Eq. \ref{eq_F1}, with $e$ = 8.0 \textmugreek m as a fitting parameter (dotted line). More importantly, Eq. \ref{eq_F1} reproduces the linearity of $F$ with the effective mass $m$ for both static and dynamic experiments, as evidenced in Figure \ref{figure3}b. The best fit is $e$ = 8.9 \textmugreek m for the static measurements and $e$ = 14.2 \textmugreek m for the dynamic experiments. Both values match the film thickness calibration with the two different frames: $e$ = 9.8 $\pm$ 1.4 \textmugreek m for the blue points and $e$ = 14.6 $\pm$ 2.3 \textmugreek m for the red points). Experimentally, the particle moves as if it were trapped in a parabolic well only determined by the mass of the film. The catenoid-like film deformation induced by the particle does not impact its dynamics, even if $h_1$ and $h_2$ are of the same order of magnitude (Figures. \ref{figure2}b and d). 

\smallskip

To understand this apparent discrepancy, we calculate the force $F_2$ due to $h_2$ only (corresponding to the limit of a particle in a weightless film). $F_2$ is found by solving the film shape equation $\Delta h_{\rm 2} = 0$ for an off-centered particle in the small deflection limit. The boundary conditions are $h_{\rm 2} = 0$ at the frame and $h_{\rm 2} = h_R$ at the equator of the particle ($h_R = f(r_{\rm m})$ is the vertical position of the centre of the particle, calculated later). In this problem, a relevant coordinate system are bipolar coordinates ($\sigma, \tau, z$), shown in Figure \ref{figure3}c. In bipolar coordinates, the circular frame is simply expressed as the iso-tau curve $\tau = \tau_L$ and the equator of the particle as $\tau = \tau_R$ (see Appendix). Solving the film equation in bipolar coordinates becomes straightforward, and gives $h_2 = h_R {(\tau - \tau_L)}/{(\tau_R - \tau_L)}$. The film deformation is associated with the surface energy $E_{\gamma} = E_0 + 2\pi\gamma h_R^2/(\tau_R - \tau_L)$. The force $F_2$ is calculated from the total energy $E_{\rm} = E_{\gamma} + mg h_R$ of the system \{particle $+$ film\}. First, the vertical equilibrium of the particle imposes $\partial E_{\rm}/\partial h_R = 0$, which gives the vertical position of the particle $h_R = - {mg(\tau_R - \tau_L)}/{(4\pi\gamma)}$. Using the expression of $h_R$, $E$ writes $E = E_0 - {(mg)^2 (\tau_R - \tau_L)}/{(8\pi\gamma)}$. When expressed in cylindrical coordinates, $E$ is anharmonic (see Appendix), but it is harmonic in the limit $r_{\rm m} \ll L$: $E \simeq E_1 + {(mg)^2 r_{\rm m}^2}/{(8\pi\gamma L^2)}$. Finally, the amplitude of the film force $F_2$ is calculated as $F_2 = \lvert - \partial E/\partial r_{\rm m} \rvert$, which writes:
\begin{align}
    F_2 = \frac{(mg)^2}{4\pi\gamma L^2}r_{\rm m} .
    \label{eq_F2}
\end{align}
\noindent $F_2$ increases quadratically with the mass $m$ of the particles, which differs from the experiments where $F \propto m$ (Figure \ref{figure3}b). This confirms that $F_1$ dominates: the key factor in the particle dynamics is the deformation of the film under its weight.

This is understood by calculating the ratio of the two forces ${F_1}/{F_2} = {\rho \pi L^2 e}/{m}$, which is the exact ratio of the masses of the film and the particle. In our experiment $m_{\rm film} \simeq 30$ mg: it is 3 to 100 times higher than the particle mass, explaining why $F_1$ dominates. By keeping the same film size, and in the limit of small, dense particles such as the ones used here ($R_{\rm b} \ll L$ and $\rho_b > \rho$) the condition $F_1 > F_2$ is almost always verified. Taking for example $\rho_b$ = 5000 kg/m$^3$, the maximum bead mass that can be held by the film is $m_{\rm max}$ = 19 mg (see section \ref{Film}), which is still smaller than the mass of the film. However, the force $F_2$ is expected to dominate when reducing the film size $L$: for a particle of mass $m = 5$ mg, it happens for $L < 1.8$ cm. Experimentally, we could not test this limit due to edge effects (the bead is attracted towards the frame by capillarity at distances of the order of 5 mm).

\section{Friction in a soap film}
\label{Friction}

\begin{figure*}
\centering
\includegraphics[width=0.9\textwidth]{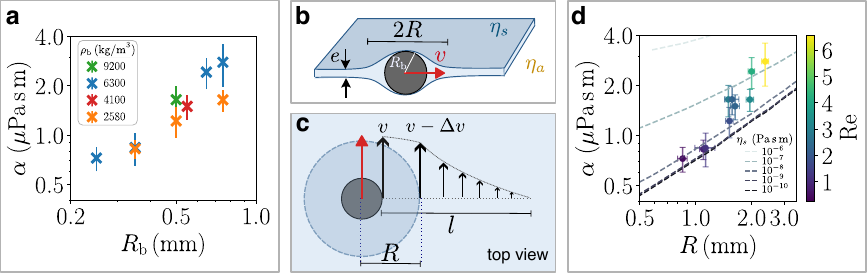}
\caption{\textbf{a.} Translational drag coefficient $\alpha$ as a function of the radius $R_{\rm b}$ of the particle, for varying marble densities. \textbf{b.} Side view of a particle immersed in a soap film. \textbf{c.} Top view. The limit of the meniscus surrounding the particle is shown with a dotted blue line. The arrows evidence the shear flow in the meniscus and the film. \textbf{d.} Comparison between model and experiments for the drag coefficient $\alpha$ as a function of the effective radius $R$ of the moving object (particle + meniscus). The color code shows the Reynolds number (in air) for each experiment. The dotted lines show the theory for varying surface viscosities $\eta_s$, from 10$^{-6}$ Pa s m (light gray) to 10$^{-10}$ Pa s m (black). The curves for 10$^{-9}$ Pa s m and 10$^{-10}$ Pa s m are almost perfectly superimposed.}
\label{figure4}
\end{figure*}

We finally focus on the friction force $\boldsymbol{F}_{\rm friction} =- \alpha \boldsymbol{\varv}$ (with $\boldsymbol{\varv}$ the particle velocity) experienced by the particle during its motion. $\alpha$ is deduced from the characteristic time $\tau$ of the exponential envelope of $x(t)$ and $y(t)$ (Figure \ref{figure1}c), using a damped harmonic oscillator model. Due to the extending meniscus around the particle (see Figure \ref{figureA1}), we expect $\alpha$ to vary with time as the effective radius of the particle increases. Experimentally, this variation is small enough not to perturb significantly exponential fit. In figure \ref{figure4}a, $\alpha$ is plotted as a function of the radius $R_b$ of the particle, the different colors correspond to different particle densities $\rho_b$. The friction coefficient is extremely small: $\alpha \simeq 1$ \textmugreek Pa\,s\,m for a millimeter-sized particle. This is of the order of the drag experienced by Leidenfrost droplets \citep{Quere:2013} and 10 times smaller than the friction of a particle floating at the surface of a bath for which $\alpha \simeq 3 \pi \eta R \simeq 10$\,\textmugreek Pa\,s\,m \citep{Danov:2000}. 

\medskip

To understand and model the friction coefficient $\alpha$, one must first consider the flow (in the film and in the air) induced by the translation at a velocity $\varv$ of a particle trapped in the film (figure \ref{figure4}b).

\smallskip

In the film, the total flow is the sum of the flow towards the bead by capillary suction (with characteristic velocity $\varv_{\rm c}$) and the flow induced by the translation of the particle. 
\noindent $\varv_{\rm c}$ is first estimated from the variation $\Delta \Omega \simeq 4 R_{\rm b}^3$ of the meniscus volume in the duration $\Delta t \simeq 30$ s of an experiment (figure \ref{figureA1}b). It gives: $\varv_{\rm c} = (\Delta \Omega/\Delta t) \times 1/(e 2\pi R) \simeq 500$~µm/s, which is more than 10 times smaller than $\varv$. The flow induced by capillary suction is therefore negligible compared to the flow induced by the motion of the bead.

\noindent Due to the mobility of the interfaces, the velocity field induced by the motion of the particle is dominated by in-plane flows, which are invariant along the $z$-direction. Figure \ref{figure4}b is a sketch from the top of the experiment, where the velocity profile $\varv = \varv_{\theta}(r)$ of the in-plane flow perpendicularly to the direction of motion of the particle is shown with black arrows. Here, the shear happens over a characteristic distance $l$ of the order of a few times $R_{\rm b}$ \citep{Stone:1998}. If we now focus on the viscous dissipation associated with the in-plane flow, a 2D apparent viscosity term appears: $\eta^{2D} = 2\eta_s + \eta e$, which is the sum of \textit{i)} the surface viscosity $\eta_s$ associated with the shearing of the 2D surfactant-rich layers at the two-liquid-air interfaces and \textit{ii)} the liquid bulk viscosity integrated over the thickness $e$ of the film, $\eta e$. The measurement of the surface viscosity of interfaces populated by soluble surfactants such as SDS has demonstrated to be particularly challenging \citep{Stevenson:2005}. However, a carefully designed experiment of \cite{Zell:2014} gave an upper bound $\eta_s < 0.01$ \textmugreek Pa s m. Here, $\eta e$ is systematically higher than this value: in a first approximation, we therefore consider that $\eta^{2D} \simeq \eta e$. The validity of this assumption, and the impact of $\eta_s$ on the friction is discussed later.

\noindent An important parameter for the friction is the presence of the meniscus surrounding the particle. It creates a region of characteristic size $R$ (see figure \ref{figureA1}a) where the film thickness is $\sim R_{\rm b}$, which is 100 times thicker than the rest of the film. To evaluate the impact of the presence of a thicker zone on the in-plane flows, we calculate the velocity $\varv-\Delta \varv$ of the fluid in the film at the boundary between the meniscus (dark blue in Figure \ref{figure4}c) and the film (light blue) using a scaling law analysis. For simplicity, we consider that the film thickness is uniform in the meniscus and suddenly decreases from $R_{\rm b}$ to $e$ at the boundary, \textit{i.e} at a distance $R - R_{\rm b} \sim R_{\rm b}$ from the equator of the particle. This induces a jump of the 2D viscosity, from $\eta^{2D} \sim \eta R_{\rm b}$ in the meniscus to $\eta^{2D}\sim \eta e$ in the film. The continuity of the surface stress $\sigma^{2D} = \eta^{2D} (\partial \varv/\partial r)$ (the shear stress integrated over the film thickness) then gives $\Delta \varv$. Using the notations of Figure \ref{figure4}b, $\sigma^{2D}$ scales as $(\eta R_{\rm b}) \Delta \varv/R_{\rm b}$ in the meniscus, and $(\eta e) (v- \Delta \varv)/l$ in the film. Equating these two terms gives $\Delta \varv \sim (e/l) \, \varv$ and $v - \Delta \varv \sim (1-e/l)\varv \simeq 0.99 \varv$, meaning that the meniscus moves \textit{with} the particle at a velocity $\varv$. In addition, in absence of a velocity gradient, there is no viscous dissipation in the meniscus. From the outside, the particle and its meniscus thus form a larger object with effective radius $R$ and effective mass $m$. Similarly to what is done for  $m$, the characteristic dimension $R$ is measured for each experiment at a time $t = 2.5 \tau$ corresponding to half of the duration of the oscillations.

\smallskip

The translation of the particle and the meniscus also induce a shear flow in the surrounding air. The Reynolds number associated with this motion writes $Re = \rho_a R \varv/\eta_a$, with respectively $\rho_a$ and $\eta_a$ the density and viscosity of air. We take the characteristic velocity $\varv$ as the average maximum velocity of the particle over the duration of the oscillations. Experimentally, $Re$ varies between 0.25 for the smallest (and slowest) particle and 6.5 for the larger one. The smallest particles thus induce a Stokes flow in the air, while for the largest particles, a visco-inertial boundary layer starts to develop. Inertial effects are expected to impact the friction: for a sphere translating in air, the friction coefficient is increased by 50~\% at $Re$ = 6 compared to the Stokes drag \citep{Munson:1995}.

\medskip 

In the following, we propose to model the friction in the (simpler) situation of $Re < 1$. In this regime, the relative contributions of the film and the surrounding air to the viscous dissipation are given by the Boussinesq number $Bo = {(\eta e + 2 \eta_s)}/{(\eta_a R)}$. For a characteristic film thickness $e$ = 10 µm, $Bo \simeq 2$: a value close to one, meaning that the bulk film and air both contribute to the friction. This configuration has been rarely considered in the literature, which is generally focused on the $Bo \ll 1$ limit (the drag of an object in a 3D fluid) and $Bo \gg 1$ (an inclusion trapped in a viscous membrane). We propose here to base our analysis on the prediction of \cite{Hughes:1981}, who solved numerically the translational drag coefficient $\alpha$ of a non-protruding cylindrical inclusion in a membrane for any arbitrary Boussinesq number. They show that, for an inclusion of radius $R$, the friction coefficient writes $\alpha = 8\pi \eta_a R \Lambda_T(Bo)$ with $\Lambda_T(Bo)$ a numerical coefficient decreasing when the Boussinesq number increases. In our system, however, the particle with radius $R_{\rm b} \gg e$ strongly protrudes from the film. To account for this, we consider in a first approximation that the friction  of the marble and meniscus in a soap film is equal to the friction of a disk of radius $R$ in a membrane surrounded by air \citep{Hughes:1981} to which we add the difference between the friction of a sphere of size $R$ in air ($\alpha = 6 \pi \eta_a R$) and that of an infinitely thin disk moving in its plane direction ($\alpha = 32\eta_a R/3$) \citep{Happel:1983}. This gives the following expression for $\alpha$: 
\begin{align}
\alpha = 8\pi \eta_a R\left[ \Lambda_T\left(\frac{\eta e + 2 \eta_s}{\eta_a R}\right) + \frac{3}{4} - \frac{4}{3\pi}\right] ,
\label{alpha}
\end{align}

In Figure \ref{figure4}d, the experimental measurement of $\alpha$ is plotted as a function of the effective radius of the moving object $R$. The colour code indicates the Reynolds number in air, varying from $Re$ = 0 (dark blue) to $Re$ = 6.5 (yellow). The dotted lines are the theoretical prediction for varying surface viscosities $\eta_s$, varied logarithmically between 10$^{-10}$ (black) and 10$^{-6}$ (light grey). For the smallest particles ($R < 2$ mm and $Re < 1$), the measured friction coefficient $\alpha$ matches the prediction of Equation \ref{alpha}, for a surface viscosity $\eta_s \leq 10^{-8}$ Pa\,s\,m. These values of the surface viscosity perfectly agree with the previous measurement of \cite{Zell:2014}, who also gave 10$^{-8}$ Pa\,s\,m as an upper boundary for the surface viscosity. 
As the Reynolds number increases, the experiments deviate from the theoretical curve: the friction coefficient $\alpha$ is higher than the one expected from a model based on purely viscous flows. An increase of $\alpha$ is coherent with inertial effects; however there is no simple way to simply model it as the friction coefficient results from a complex interplay between the flow in the film and in air \citep{Hughes:1981}.

\smallskip


\section{Conclusion}

To conclude, we measured and modelled the forces that apply to a millimeter-sized particle trapped in a soap film, by focusing both on the static position of the bead and its motion. We show that the gravitational distortion of the film under its weight -- of only a few hundreds of micrometers -- is the key to understand the bead dynamics. Indeed, in our experiment, the component of the force due to the film weight systematically dominates over the force induced by the weight of the particle. This is expected as long as the mass of the film is larger than the mass of the particle, which happens for solid particles of size $R \ll L$. Counter-intuitively, decreasing the size of the film while keeping the same ratio between the particle radius and the film width $L/R \simeq$ 100 will only increase the relative importance of the film weight. For example, for a 10 \textmugreek m particle trapped on an 1 mm$^2$ horizontal film, $F_1/F_2 >10$ even for very thin films $e \simeq 100$ nm.

In a second part, we focus on the drag force experienced by the particle, and we propose a model in the limit of low Reynolds number in air ($Re < 1$). In this regime, the particle is submitted to the viscous friction of air and to that of the film, with an almost equal contribution. A simplified model based on the work of \cite{Hughes:1981} matches exactly our measurements for the smallest particles. A deviation is observed as the Reynolds number (and the size of the particle) increase, which we interpret as the effect of the inertia of the air. Interestingly, the match between the theory and the experiments is valid for all surface viscosities $\eta_s \leq 10^{-8}$ Pa\,s\,m: a mismatch of at least 50\% is expected for $\eta_s \geq 10^{-7}$ Pa\,s\,m. This suggests that our experiment should be sensitive to surface viscosities as low as $10^{-8}$~Pa\,s\,m, which is as sensitive as the most precise interfacial rheology setups, for example controlled microrheological probes at the surface of a bath \citep{Zell:2014}. We expect our model to remain valid when reducing the size of the probe, which opens interesting perspectives on the possibility of using small particles as probes to explore in-situ the rheological properties of soap film.

\medskip

\noindent \textbf{Acknowledgments.} The authors thank Devaraj van der Meer for the initial discussions on the topic, and Marwane Taoufiki and Hugo Delrieux for preliminary experiments.

\medskip

\noindent \textbf{Declaration of Interests.} The authors report no conflict of interest.

\appendix
\counterwithin{figure}{section}

\section{Meniscus characterization}

A particle trapped in a soap film is surrounded by a meniscus which extends with time by capillary suction. Figure \ref{figureA1} characterises the shape of the meniscus, and present its growth dynamics in a typical experiment. The presence of the meniscus increases the effective mass $m$ of the moving object, which is taken into account in the calculation of the forces.

\bigskip

\begin{center}
\begin{minipage}{\textwidth}
    \centering
    \includegraphics[width=0.99\textwidth]{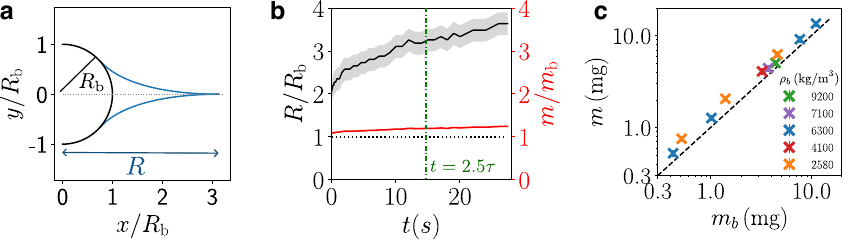}
    \captionof{figure}{\textbf{a.} Shape of a meniscus with an extension $R = 3R_{\rm b}$ around a particle, calculated based on \cite{Orr:1975}. The position of the interface is normalized by the particle radius $R_b$. \textbf{b.} Normalized extension $R/R_{\rm b}$ of the meniscus (black line) and mass $m/m_{\rm b}$ (red line) as a function of time $t$, for a particle of radius $R_b = 0.5$ mm and mass $m_b$ = 3.7 mg. The black dotted line evidences the limit $R = R_{\rm b}$ and $m = m_{\rm b}$. The vertical green line evidences the time $t = 2.5 \tau$ at which the characteristic size $R$ and mass $m$ are measured. $t = 2.5 \tau$ is half of the duration of the oscillations. \textbf{c.} Total mass $m$ of the particle and its meniscus, measured at a time $t = 2.5 \tau$, as a function of the mass $m_{\rm b}$ of the particle alone. The dotted line shows the limit $m = m_{\rm b}$.}
   \label{figureA1}
\end{minipage}
\end{center}

\section{Calculation of the film-mediated force in bipolar coordinates}

We want to determine the energy of the system consisting of a particle of radius $R_{\rm b}$ off-centered by $r_m$ in a circular soap film of radius $L$. In bipolar coordinates, the iso-$\tau$ curves are circles of radius $c/|\sinh\tau|$, centred in ($x = c\coth\tau$, $y = 0$), with $c$ a constant. One needs to find the constant $c$ so that the frame position is simply expressed as $\tau = \tau_L$ and the edge of the particle is $\tau = \tau_R$ (as shown in Figure \ref{figure3}c). We can restrict $\tau_L$ and $\tau_R$ to be positive, and by construction of the bipolar coordinates, $\tau_R > \tau_L$. The relation between $\tau_L$, $\tau_R$, $c$ and the dimensions of the problem $L$, $R_{\rm b}$ and $r_m$ is solved by inverting the three relations: $r_m = c(\coth{\tau_L} - \coth{\tau_R})$ $(r_m > 0)$, $L = c/\sinh{\tau_L}$ and $R_{\rm b} = c/\sinh{\tau_R}$. In particular:
\begin{align}
    \cosh{\tau_L} = \frac{L^2 - R_{\rm b}^2 + r_m^2}{2Lr_m} \text{ and } \cosh{\tau_R} = \frac{L^2 - R_{\rm b}^2 - r_m^2}{2Rr_m} .
    \label{tau_R}
\end{align}

\noindent In bipolar coordinates, $\Delta h = (\cosh\tau - \cos\sigma)^2 (\partial^2 h/\partial\sigma^2 + \partial^2 h/\partial\tau^2)/c^2$ \citep{Happel:1983}. The film shape equation $\Delta h_2 = 0$ is then solved with the boundary conditions $h(\tau_L) = 0$ and $h(\tau_R) = h_R$, giving a simple solution: $h_2 = h_R (\tau - \tau_L)/{(\tau_R - \tau_L)}$. In the small deflection limit, the surface energy associated with the film deformation is $E_{\gamma} \simeq E_0 + \gamma \iint (\nabla h)^2 dS$. A small surface element is $dS= {c^2 \mathrm{d}\sigma\,\mathrm{d}\tau}/{(\cosh{\tau} - \cos{\sigma})^2}  $ while $\mathbf{\nabla} h = (\cosh{\tau}-\cos{\sigma}) ( \mathbf{e}_{\sigma} {\partial h}/{\partial \sigma} + \mathbf{e}_{\tau} {\partial h}/{\partial \tau} )/c$. Using the expression of $h_2$, the film energy is:
\begin{align}
E_{\gamma} = E_0 + \gamma \int_0^{2\pi} d\sigma \int_{\tau_L}^{\tau_R} \left(\frac{\partial h_2}{\partial \tau}\right)^2 d\tau = E_0 + 2\pi\gamma \frac{h_R^2}{(\tau_R - \tau_L)}.
\end{align}

The total energy $E$ of the film and particle is $E = E_{\gamma} + mgh_R$. The vertical equilibrium of the particle imposes $\partial E_{\rm}/\partial h_R = 0$, which gives $h_R = -{mg(\tau_R - \tau_L)}/{4\pi\gamma}$ and $E = E_0 - {(mg)^2 (\tau_R - \tau_L)}/{(8\pi\gamma)}$. Using Eq. \ref{tau_R}, and in the limit $R_{\rm b} \ll L$, $E$ finally writes:
\begin{align}
    E = E_0 - \frac{(mg)^2}{8\pi\gamma}\left[ \ln \frac{L}{R_{\rm b}} + \ln \left(1-\frac{r_m^2}{L^2} + \sqrt{1-2\frac{r_m^2}{L^2}}\right) - \ln \left(1+\frac{r_m^2}{L^2} + \sqrt{1-2\frac{r_m^2}{L^2}}\right) \right] .
\end{align}




\bibliographystyle{jfm}
\bibliography{jfm.bib}

\end{document}